\documentclass[aps,reprint,showpacs,showkeys,floatfix,longbibliography,superscriptaddress,prstper]{revtex4-1}

\usepackage{graphicx}
\usepackage{dcolumn}
\usepackage{amsmath}
\usepackage{amssymb}
\usepackage{epstopdf}
\usepackage{rotating}
\usepackage{color}

\bibpunct{[}{]}{,}{n}{}{}

\newcommand{\eval}[1]{\textrm{\hspace{0.5mm}\rule[-3mm]{0.25mm}{8mm}}_{_{ \hspace{0.25mm} #1 }}\textrm{\hspace{-2.5mm}}}

\newcommand{\bft}{\textit{Boltzmann Factor} tutorial}
\newcommand{\BF}{Boltzmann factor}
\newcommand{\Z}{canonical partition function}
\newcommand{\D }{\mathrm{d}}

\newcommand{\un}[1]{\mathrm{~#1}}

\begin{document}

\title{Student Understanding of the Boltzmann Factor}


\author{Trevor I. Smith}\affiliation{Department of Physics \& Astronomy and Department of Teacher Education, Rowan University, Glassboro, NJ 08028}

\author{Donald B. Mountcastle}\affiliation{Department of Physics \& Astronomy, University of Maine, Orono, ME 04469} 

\author{John R. Thompson}\affiliation{Department of Physics \& Astronomy, University of Maine, Orono, ME 04469}\affiliation{Maine Center for Research in STEM Education, University of Maine, Orono, ME 04469}

\begin{abstract}
We present results of our investigation into student understanding of the physical significance and utility of the \BF\ in several simple models. We identify various justifications, both correct and incorrect, that students use when answering written questions that require application of the \BF.  Results from written data as well as teaching interviews suggest that many students can neither recognize situations in which the \BF\ is applicable, nor articulate the physical significance of the \BF\ as an expression for multiplicity, a fundamental quantity of statistical mechanics.  The specific student difficulties seen in the written data led us to develop a guided-inquiry tutorial activity, centered around the derivation of the \BF, for use in undergraduate statistical mechanics courses. We report on the development process of our tutorial, including data from teaching interviews and classroom observations on student discussions about the \BF\ and its derivation during the tutorial development process. This additional information informed modifications that improved students' abilities to complete the tutorial during the allowed class time without sacrificing the effectiveness as we have measured it. These data also show an increase in students' appreciation of the origin and significance of the \BF\ during the student discussions. Our findings provide evidence that working in groups to better understand the physical origins of the canonical probability distribution helps students gain a better understanding of when the \BF\ is applicable and how to use it appropriately in answering relevant questions.  
\end{abstract}

\maketitle

\section{Introduction}

The study of student understanding of advanced topics is becoming increasingly prevalent in physics education research \cite{Singh2001,Singh2006,Thompson2006,Pollock2007,Bucy2007a,Chasteen2008,Black2009,Loverude2009,Smith2009,Hayes2010,Loverude2010,Mason2010,Smith2010,Pepper2012,Smith2013,Zwickl2013,Loverude2015a,Loverude2015b}. Investigating upper-division undergraduate students provides a snapshot of the intellectual journey from novice introductory student to expert physicist that may reveal key components of this transition \cite{Bing2012}. Moreover, the National Research Council has recently emphasized the need for more study of advanced undergraduate education in many science disciplines \cite{NRCDBER2012}. As part of a broader study on student learning in thermal physics, we have investigated student understanding of the Boltzmann factor with the goal of developing instructional strategies to improve that understanding. 

Statistical mechanics provides a mechanism for understanding the emergence of macroscopic phenomena from the collective properties of individual microscopic systems; as such, it is a cornerstone of contemporary physics. However, due to its complexity and sophistication, students do not typically encounter statistical mechanics until late in their undergraduate (or even graduate) studies, and comparatively little research has been done to document student difficulties and successes in this field \cite{Bucy2006,Mountcastle2007,Loverude2015a,Loverude2015b,Smith2010,Smith2013}. This work showed that even after instruction students often struggle to distinguish \textit{micro}states of a system from \textit{macro}states and to appropriately relate the two. The fundamental assumption of statistical mechanics states that all accessible microstates of a system (microscopic arrangements of a system's particles in phase space) are equally probable \cite{Schroeder2000}. Microstates that share common macroscopic properties (system volume, internal energy, etc.)\ may be grouped into measurable macrostates. The probability of finding the system in a particular macrostate, $\mathcal{M}_i$, is determined by the number of microstates corresponding to that macrostate, i.e., the multiplicity, $\omega_i$, normalized by the total number of microstates:
\begin{equation}
P(\mathcal{M}_i)=\frac{\omega_i}{\sum\limits_j\omega_j}.\label{probmult}
\end{equation}
Much of the intellectual effort of statistical mechanics is spent defining the relevant properties of the microstates and macrostates and determining the multiplicity given the macroscopic properties of the system \footnote{It is not typically possible to determine the precise microstate of the system at any given time, so relevant quantities must be related to measurable (macroscopic) properties.}.

Loverude reports that many students have difficulty distinguishing microstates and macrostates in the context of binary systems \cite{Loverude2015a}. In one question he asked students, after flipping six coins, if the probability of getting five heads was more than, less than, or equal to the probability of getting six heads; about 20\% of the students incorrectly stated that the probabilities were the same, often claiming, ``all probabilities have equal occurrences,'' which is true for microstates but not for macrostates \cite[p.\ 190]{Loverude2015a}. In another question, students had to compare the probabilities of a six-child family having two different sequences of boys and girls (GBGBBG vs.\ BGBBBB). Over one third of students incorrectly stated that the second sequence was less probable because families are more likely to have equal numbers of boys and girls rather than only one girl out of six, thus connecting the probabilities of a macrostate (the relative number of boys and girls) to an individual microstate (a specific birth sequence).

Loverude also provides evidence that students struggle to distinguish microstates from macrostates, especially in the context of interacting systems. In the context of Einstein's model for a solid lattice structure, Loverude asked students to determine the most likely energy distribution between two lattices of different sizes \cite{Loverude2015b}. About 40\% of students incorrectly stated that the most probable macrostate is the one in which each solid has the same amount of energy and disregarded the number of oscillators within each lattice. Loverude also reports that students often add the multiplicities of interacting Einstein solids to determine the total multiplicity rather than appropriately multiplying them \cite{Loverude2015b}. 

A key aspect of equilibrium statistical mechanics is that, when dealing with large systems ($\sim10^{23}$ particles), the most likely state of the system is overwhelmingly the most probable. This result is due to the fact that the statistical spread of the macrostate probability distribution tends to decrease as $\sigma\propto N^{-1/2}$ where $\sigma$ is the standard deviation and $N$ is the number of particles in the system. When $N$ is large, nearly 100\% of all microstates exist within a range of macrostates that are virtually indistinguishable from each other; i.e., within the limits of measurable uncertainty. This single most likely ``system state'' is the equilibrium state (with microscopic fluctuations) of the macroscopic thermodynamic system \footnote{The single most probable macrostate actually has a very low probability when $N$ is large. The ``overwhelming'' probability comes solely from the fact that the statistical variance is so small as to make neighboring macrostates indistinguishable.}.

Mountcastle, Bucy, and Thompson studied students' understanding of probability distributions by asking them to determine the most probable number of ``heads'' when flipping $N$ coins as well the uncertainty in this value (reported as $a\pm\Delta a$) \cite{Mountcastle2007}. About a third of students incorrectly indicated that the relative uncertainty remains constant as $N$ increases, e.g., $\Delta a/a=15\%$ for all cases, and about 20\% stated that the uncertainty covers the entire range of possible values (the most probable result is $N/2\pm N/2$). However, students readily recognized that performing additional measurements would reduce the uncertainty of the mean; e.g., using more rain gauges to measure amount of rainfall \cite{Mountcastle2007}. Further investigation showed that students have difficulty reconciling the ``overwhelmingly probable'' equilibrium state with calculations and graphs showing that the probability of the single most likely macrostate actually decreases with increasing $N$: $P_{\textrm{max}}=N!/\left[2^N\left(N/2\right)!\left(N/2\right)!\right]$ for the binomial distribution. Some students took this idea to the extreme on the coin toss question by stating that the most probable result of flipping $6\times10^{23}$ coins is $3\times10^{23}\pm1$ heads. The distinction between a single discrete macrostate and an equilibrium thermodynamic ``state'' (consisting of a range of virtually indistinguishable macrostates) is subtle and requires careful attention by both students and instructors. These results, along with Loverude's \cite{Loverude2015b}, provide the foundation for studies into students' understanding of the statistical treatment of thermodynamic systems where states are defined by continuous (rather than discrete) quantities.

The canonical probability distribution defined by the \BF\ has been described as ``the quintessential expression of the statistical mechanical approach'' \cite[p.\ 109]{Baierlein1999} and ``the most powerful tool in all of statistical mechanics'' \cite[p.\ 220]{Schroeder2000}. By knowing the possible microscopic energy eigenstates, one may deduce the thermodynamic equilibrium properties of any system at constant temperature, including average internal energy, free energy, entropy, pressure, heat capacity, etc. This connection between microscopic and macroscopic properties is known to be difficult in multiple contexts in physics \cite{Eylon1990,Thacker1999,Kautz2005a,Kautz2005b,Loverude2015a,Loverude2015b} as well as chemistry \cite{Cooper2013}.  As the core of statistical mechanics is this micro-macro connection, this topic is an optimal context for an investigation of this nature.  Our investigation of student understanding of the \BF\ provides additional information about difficulties students have with this connection; these results have implications for studies of more complex systems and topics. 

In this paper we present results of our investigation into student understanding of the physical significance and utility of the \BF\ in several simple models. We identify various justifications, both correct and incorrect, that students use when answering written questions that require application of the \BF.  Results from written data as well as teaching interviews suggest that many students can neither recognize situations in which the \BF\ is applicable, nor articulate the physical significance of the \BF\ as an expression for multiplicity, a fundamental quantity of statistical mechanics.  The specific student difficulties seen in the written data led us to develop a guided-inquiry tutorial activity, centered around the derivation of the \BF, for use in an undergraduate statistical mechanics course. We report on the development process of our tutorial, including data from teaching interviews and classroom observations on student discussions about the \BF\ and its derivation during the tutorial development process. This additional information informed modifications that improved students' abilities to complete the tutorial during the allowed class time without sacrificing the effectiveness as we have measured it. Our findings provide evidence that working in groups to better understand the physical origins of the canonical probability distribution helps students gain a better understanding of when the \BF\ is applicable and how to use it appropriately in answering relevant questions.

\section{The Physics of the Boltzmann Factor}
\label{sec:bf-physics}
Before discussing our research on student understanding of the \BF, it is useful to provide an overview of the physics (and mathematics) of the \BF\ and the \Z. The particular derivation of the \BF\ and the \Z\ through which students are guided in the tutorial is included in the Appendix. 

\begin{figure}[tb]
\includegraphics[width=1.5in]{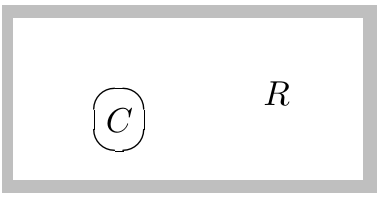}
\caption{Sample system for the \BF\ instructional sequence. An isolated container of an ideal gas is separated into a small system ($C\,$) and a large reservoir ($R\,$). The label ``$C\,$'' is used to avoid confusion with entropy.}
\label{sys-res}
\end{figure}

The underlying assumption of the canonical ensemble is that the thermodynamic system has a fixed equilibrium temperature, a fixed number of particles, and may exchange energy with its surroundings. A standard model for the canonical ensemble is a very small system in equilibrium with a large thermal energy reservoir (free to exchange energy but not particles, see Fig.\ \ref{sys-res}). The Boltzmann factor is a mathematical expression for the probability that a system in equilibrium at a fixed temperature is in a particular energy state,
\begin{equation}
P(\psi_j)\propto e^{-E_j/kT},\label{bf}
\end{equation}
where $\psi_j$ denotes the microstate with a particular energy $E_j$, $k$ is Boltzmann's constant, and $T$ is the temperature of the system \footnote{The \BF\ may also be used to analyze systems with changing temperatures provided that the change is slow enough that a quasistatic approximation is valid.}. The decaying exponential form of the \BF\ results from an expression of the multiplicity of the reservoir derived from Boltzmann's equation, 
\begin{equation}
S=k\ln(\omega).\label{S}
\end{equation} 
As the energy of the system decreases, the energy (and multiplicity) of the reservoir increases in such a way that the total probability increases (see the Appendix for full details).

The \Z\ ($Z$) is the result of the normalization constraint that the sum of probabilities [$P(\psi_j)$] over all $j$ must be unity:
\begin{align}
\sum\limits_{j}P(\psi_j)&=\sum\limits_{j}\frac{e^{-E_j/kT}}{Z}=1\label{prob}\\
Z&=\sum\limits_{i}e^{-E_i/kT},\label{Z}
\end{align}
where $Z$ depends on temperature, but is independent of the energy value $E_j$ \footnote{In this expression for $Z$ one cannot assume that all values of $E_i$ are unique; duplicate terms appear in the summation to account for all degenerate macrostates.}. One may also express the partition function in terms of the energy of a macrostate, $E$:
\begin{align}
Z&=\int\limits_{All~E}D(E)\,e^{-E/kT}\,\mathrm{d}E,\label{Z-cont}
\end{align}
where the density of states function, $D(E)$, accounts for the degeneracy (or multiplicity) of the macrostate. In this way the \Z\ is equally valid for systems with discrete energy microstates, as in Eq.\ \eqref{Z}, and those with continuous energy distributions, as in Eq.\ \eqref{Z-cont}.

The \Z\ can be used to express equilibrium (macroscopic) thermodynamic quantities.  For example, the Helmholtz free energy of a system may be written as a function of $Z$,
\begin{equation}
F=-kT\ln(Z);\label{F}
\end{equation}
derivatives of $F$ yield information about the system's entropy, pressure, magnetization, and many other thermodynamic variables. Moreover, the average energy of a system, $\langle{E}\rangle$, may be expressed as a derivative of the natural logarithm of $Z$. Due to these connections, Schroeder refers to the canonical probability distribution as ``the most useful formula in all of statistical mechanics'' \cite[p.\ 223]{Schroeder2000}. The \Z\ and the \BF\ are cornerstones of statistical mechanics, and a thorough understanding of \textit{when} and \textit{how} they are useful (when examining an equilibrium system at constant temperature) is essential for study in the field.

\section{Student Understanding of the Origin and Utility of the Boltzmann Factor}
\label{sec:diffs-bf}
We gathered data in several different forms to study student understanding of the \BF\ from multiple perspectives. In one investigation, we gave students an ungraded written survey to determine whether or not they use the \BF\ in an appropriate context. Additionally, we conducted teaching experiments with several students as well as classroom observations to assess their understanding of the physical origin and significance of the mathematical expression of the \BF. Our results indicate that students often do not use the \BF\ in appropriate contexts, instead using vague notions of lower energies having higher probabilities 
to make conclusions about the ratios of these probabilities
. Moreover, we find that students may not recognize the physical significance of the \BF, even after having memorized its mathematical derivation.

\begin{figure*}[tb]
\begin{center}
\includegraphics[width=6.5in]{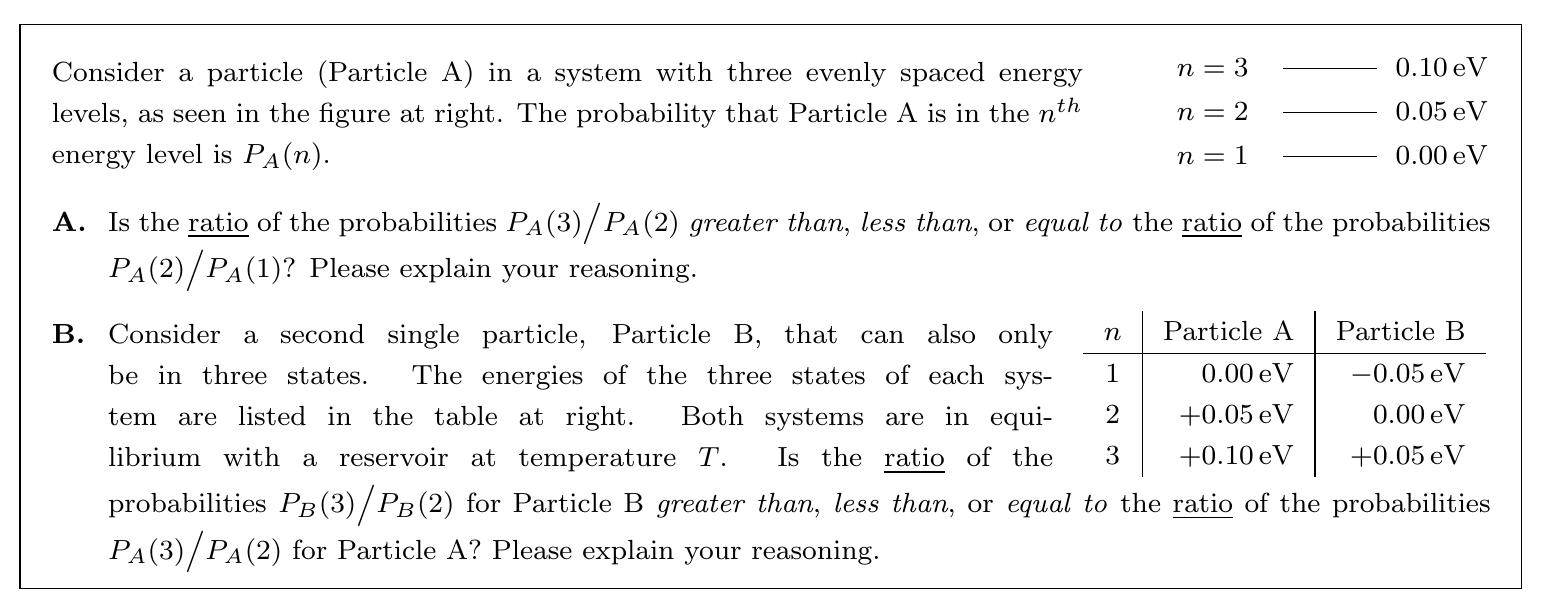}
\end{center}
\caption{Probability Ratios Question (PRQ): given as an ungraded survey before tutorial instruction.}
\label{prq}
\end{figure*}

\subsection{Student use of the Boltzmann factor in appropriate contexts} 
\label{sec:prq-pre}
One desired result of teaching students about the \BF\ is that they will recognize applicable situations and use it appropriately to make claims about probabilities of the occupation of specific energy states. The probability ratios question (PRQ, shown in Fig.\ \ref{prq}) probes their ability to do this. The correct solution to the PRQ requires students to recognize three pieces of information:
\begin{itemize}
\item{The probability of a single particle being in each of three energy states is proportional to the \BF\ for each state}
\item{A ratio of exponential functions is the exponential of the difference of their exponents}
\item{The differences in energies between adjacent states are the same for each particle ($\Delta E_{n,n-1}\nobreak=\nobreak0.05\,\mathrm{eV}$)}
\end{itemize}
The first two items indicate that each ratio of probabilities is an exponential function of the energy difference between the two states. The third item reveals that both pairs of ratios in the PRQ are equal \footnote{In general a student's understanding of each of these three pieces of information cannot be measured independently, but a correct solution should show evidence that the student recognizes all of them.}. Students were also considered to have given a correct explanation to part B of the PRQ if they stated that the two ratios were equal because the only difference between the two particles is the energy of the ground state.

\subsubsection{Recognizing the need for the Boltzmann factor} 
\label{sec:pre-bf}
The PRQ was administered to students in an upper-division statistical mechanics course at a land-grant research university in the northeastern United States (School 1); data were collected from seven successive classes ($N=50$). Students at School 1 are typically senior undergraduates who have competed studies in classical mechanics, electrodynamics, and quantum mechanics. The PRQ was also administered once in a single-semester upper-division thermal physics course at a comprehensive public university in the western United States (School 2, $N=32$). Students at School 2 are typically junior undergraduates who have completed studies in modern physics and classical mechanics. The PRQ was administered at both schools immediately before students participated in guided-inquiry activities regarding the \BF\ and the \Z\ (our \bft, see Sec.\ \ref{sec:bf-tut}). At School 1 the activities were used after lecture instruction, and the PRQ was given after lectures. At School 2 the activities were used in place of lecture instruction, and the PRQ was given before instruction to establish a baseline for students' understanding before the tutorial.

\begin{figure*}[bt]
\begin{center}

	\includegraphics[width=3.4in]{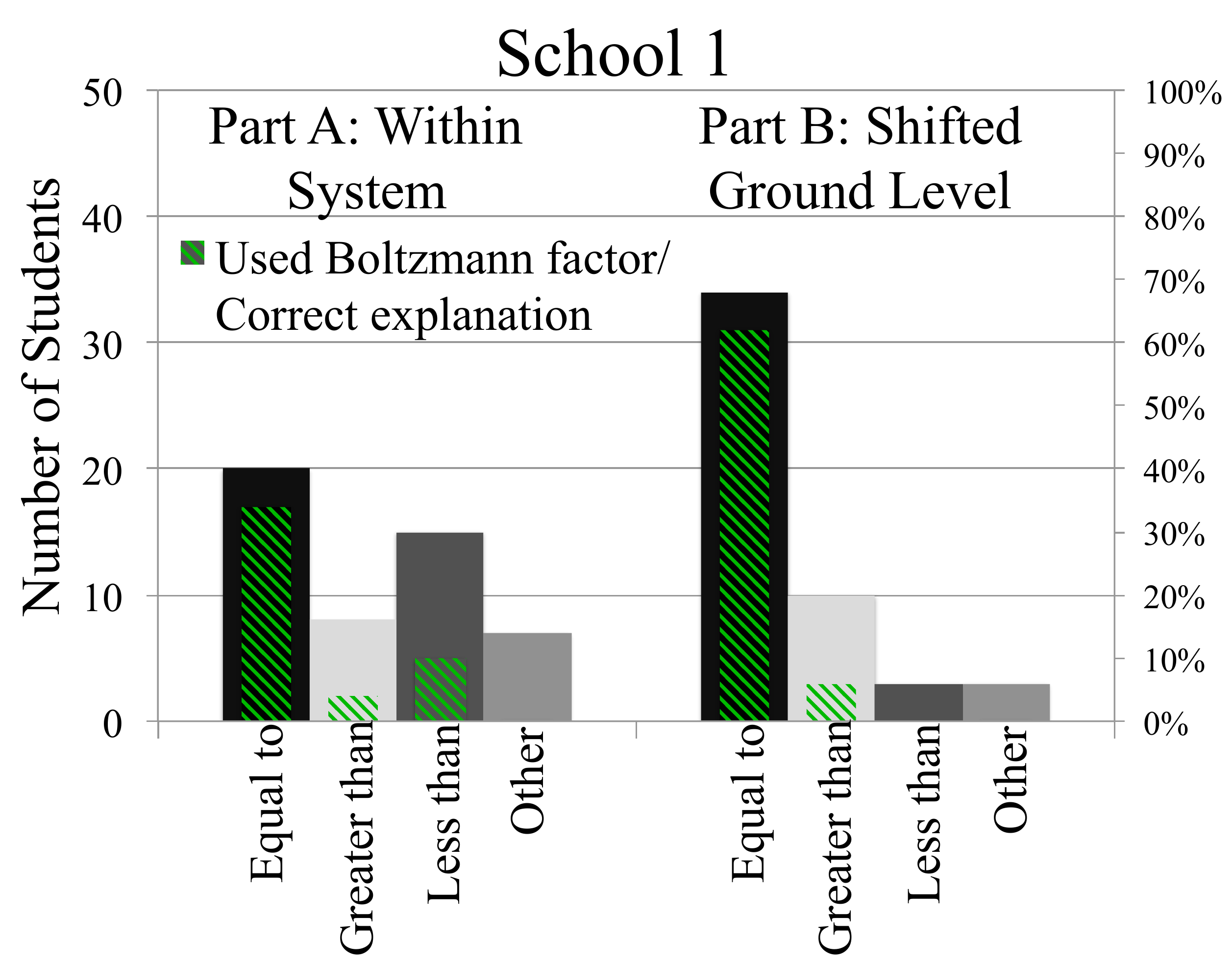}
	\hspace{0.15in}
	\includegraphics[width=3.4in]{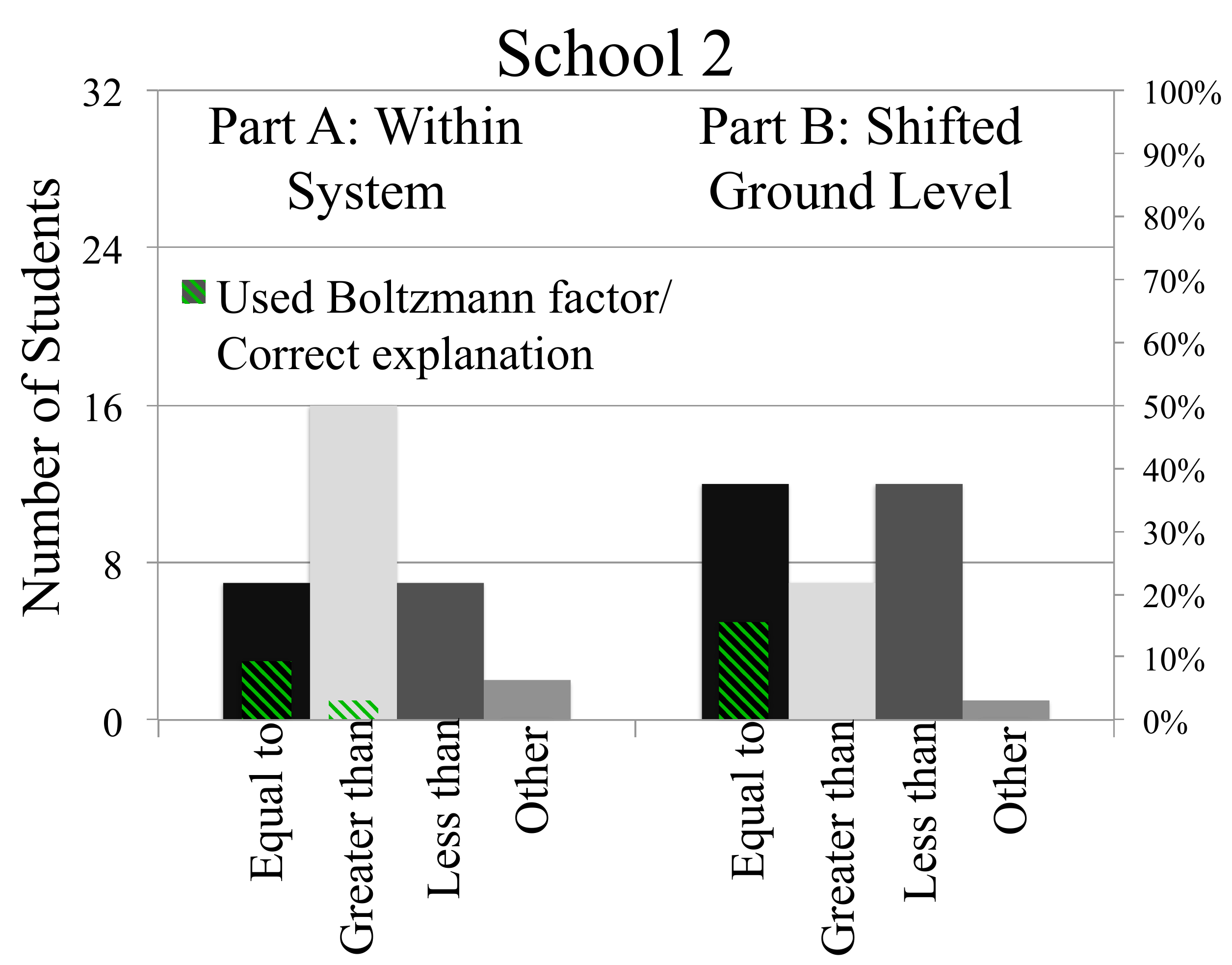}

(a)\hspace{3.55in}(b)
\end{center}
\caption{PRQ pretutorial results. a) School 1: after lecture instruction on the \BF\ over seven years ($N\nobreak=\nobreak50$), b) School 2: before any instruction on the \BF\ in one year ($N\nobreak=\nobreak32$). The green diagonal stripes indicate the students who used the Boltzmann factor or stated that the energy of the ground state was irrelevant (in part B) to obtain their chosen answers. Students in the ``Other'' column often provided no explicit answer or stated that there was not enough information to determine the answer. Only 24 students from School 1 and four students from School 2 used the Boltzmann factor on both parts.}
\label{bf-pre}
\end{figure*}

Student responses to the PRQ were coded in two ways: first by the response given (equal to, greater than, less than, or other), second by whether or not the \BF\ was used. Figure \ref{bf-pre}a shows the response frequencies for the entire seven-year data corpus from School 1, and Fig.\ \ref{bf-pre}b shows the response frequencies from School 2. Green diagonal stripes indicate the students who used the Boltzmann factor or stated that the energy of the ground state was irrelevant (in part B) to obtain their chosen answers; these students are considered to have used correct explanations regardless of which answer they chose. \footnote{This color scheme is used for all presentations of response frequency data for probability ratio tasks.}.

We used a grounded theory approach to analyze students' explanations for their responses; the entire data corpus was examined for common trends, yielding categories defined by the data, and all data were reexamined to group them into those defined categories \cite{Behrens2004,Strauss1990}. One goal of our analysis was to focus on describing rather than interpreting students' explanations while defining the categories. In this way our analysis stays as true to the data as possible by limiting researcher biases and interpretations. This is consistent with Heron's identification of specific difficulties \cite{Heron2003}.

The data represented in Fig.\ \ref{bf-pre} suggest two questions: 1) What is the prevalence of invocation of the \BF, regardless of the correctness of the response? and 2) How do students justify their answers if they do \textit{not} apply the \BF? To answer the first of these questions, the data show four categories of responses:
\begin{itemize}
\item Correct response (equal to) using the \BF\ (or stating that the energy of the ground state was irrelevant in part B) 
\item Correct response without using the \BF
\item Incorrect response using the \BF
\item Incorrect response without using the \BF
\end{itemize}

This coding scheme enables highlighting of the number of students who are and are not invoking the \BF\ to answer the PRQ. A natural question associated with this coding scheme is, how might someone invoke the \BF\ but arrive at an incorrect response? One route is to make a computational error. On the other hand, one could compare the wrong ratios, but do so correctly using the \BF. Data also indicate that some students imposed degeneracy terms when using the \BF\ to answer the PRQ. In coding responses, a student who wrote that probability is related to a decaying exponential of the energy was coded as using the \BF\ independent of the final answer obtained. Using the \BF\ and stating that the energy of the ground state is irrelevant were grouped together because both are correct physical justifications for concluding that the ratios of probabilities in part B of the PRQ are equal.

\begin{table}[b]
\caption{Results from the PRQ pretest at both School 1 ($N=50$) and School 2 ($N=32$). Students are grouped by whether or not they gave the correct answer (Equal to) and whether or not they used the Boltzmann factor (or stated that the energy of the ground state was irrelevant in part B).}
\begin{center}
\begin{tabular}{rrrrrrrr}
\hline\hline
&&&Part A&&&Part B&\\
&&Correct&Incorrect&Total&Correct&Incorrect&Total\\
\hline
&Used Bf&34\%~~~&14\%~~~~&48\%~&62\%~~~&6\%~~~~&68\%~\\
\raisebox{1.5ex}[0pt]{S1}&No Bf&6\%~~~&46\%~~~~&52\%~&6\%~~~~&26\%~~~~&32\%~\\
&Used Bf&9\%~~~&3\%~~~~&12\%~&16\%~~~&0\%~~~~&16\%~\\
\raisebox{1.5ex}[0pt]{S2}&No Bf&13\%~~~&75\%~~~~&88\%~&22\%~~~&63\%~~~~&84\%~\\
\hline
\end{tabular}
\end{center}
\label{prq-pre-tab}
\end{table}%

Table \ref{prq-pre-tab} shows the percentages of students who occupy each of the four response categories at each school for both parts of the PRQ. From the data shown in Fig.\ \ref{bf-pre} and Table \ref{prq-pre-tab} it is clear that the distribution of responses is different at the two schools. A Fisher's exact test showed this to be true ($p=0.008$ for part A, $p<0.001$ for part B) \cite{Everitt1999,Denenberg1976,Coladarci2004}. Another Fisher's exact test showed that students at School 1 are using the \BF\ on the PRQ pretest more than students at School 2 ($p<0.001$ on both parts) \footnote{The threshold for significance for all statistical tests was set at $\alpha=0.05$.}. This is not surprising given that students at School 1 had received lecture instruction on the \BF, while students at School 2 had not. On the other hand, only 48\% of students at School 1 used the \BF\ on part A and only 68\% did so on part B, indicating that lecture instruction alone was not sufficient for all students to gain a robust understanding of when and how to use the \BF.

The most common incorrect response at School 1 for both parts of the PRQ is that $P(0.10\un{eV})/P(0.05\un{eV})\nobreak<\nobreak{P(0.05\un{eV})}/P(0.00\un{eV})$ (``less than'' for part A and ``greater than'' for part B; see Table \ref{bf-pre-diff}). These answers are considered consistent because the second and third energy levels in particle B have the same numerical values as the first and second energy levels in particle A, respectively. A Fisher's exact test shows the distribution of ``less than'' and ``greater than'' responses from School 1 to be significantly different for part A as compared to part B ($p\nobreak=\nobreak0.035$). However, the data from School 2 show the exact opposite trend: more students answer ``greater than'' for part A and ``less than'' for part B (see Fig.\ \ref{bf-pre}b and Table \ref{bf-pre-diff}). A Fisher's exact test shows that this difference at School 2 approaches significance ($p=0.061$). Additional tests show that the results from School 2 are significantly different from those at School 1 ($p=0.038$ for part A and $p=0.036$ for part B).

\begin{table}[b]
\caption{Pretest Response Comparison: ``greater than'' vs.\ ``less than.'' Numbers shown indicate the percentage of \textit{incorrect} responses at each of the two schools. This is necessary because significantly more students answered the PRQ correctly at School 1 than at School 2. Only by looking at the percentages of incorrect responses can meaningful comparisons be made.}
\begin{center}
\begin{tabular}{ccccc}
\\
\hline\hline
&\multicolumn{2}{c}{Part A}&\multicolumn{2}{c}{Part B}\\
&greater than&less than&greater than&less than\\
\hline
School 1&23\%&38\%&44\%&19\%\\
School 2&54\%&25\%&26\%&44\%\\
\hline
\end{tabular}
\end{center}
\label{bf-pre-diff}
\end{table}

Figure \ref{bf-pre}a also shows that students at School 1 are more likely to answer part B correctly (which discusses an effective shift in the ground state energy of a system) than part A (comparing two different sets of probabilities for states within the same system), with only 34\% using the \BF\ to obtain the correct response on part A compared to 64\% providing a correct explanation on part B (statistically significant, $p=0.035$). One student at School 1 justified his response for part B in stating that, ``\dots it does not matter what the `baseline' is, just the amount of energy added.'' This higher performance on part B could be a result of our coding scheme in that explanations involving comments about the arbitrariness of the ground state energy were considered correct for part B regardless of the student's response to part A. This phenomenon is not significantly observed at School 2 (see Fig.\ \ref{bf-pre}b, Fisher's exact test yields $p=0.45$).

\subsubsection{Incorrect reasoning about probability ratios}
\label{sec:pre-reason}
The justifications students used to support their final answers were sorted into several categories developed using a grounded theory approach. At School 1, 24 students (out of 50) used the \BF\ within their explanation of their answers on the PRQ; only four out of 32 students at School 2 used the \BF. Of the remaining students at each school, roughly half (15 out of 26 at School 1 and 13 out of 28 at School 2) used a ranking of probabilities as their primary justification; e.g., $P_A(1)>P_A(2)>P_A(3)$. An additional five students at School 2 stated that the lowest energy is most probable but did not make claims about the relative probabilities of energy states 2 and 3. Using probability ranking, either explicit or implied, is the most common incorrect justification at both schools, and no students provided a physical explanation for why the probabilities of the various energy levels would be ranked as they claimed. 

Of the students who ranked the probabilities to justify their answers, eight students at School 1 and seven at School 2 made claims about the relative difference in probability between states 1 and 2 and between states 2 and 3. Some claims were made in sentence form, e.g., ``\dots\ it is more likely that the system will have \underline{less energy} so the difference between [states] 3 \& 2 is less than [between states] 2 and 1'' (student's emphasis); other claims took the form of a mathematical expression, e.g., ``$P_A(1)-P_A(2)\nobreak>P_A(2)-P_A(3)$.'' Both of these statements imply the idea that $P_A(1)\gg P_A(2)>P_A(3)$. All seven students at School 2 used this idea to claim that $P_A(3)/P_A(2)>P_A(2)/P_A(1)$. However, the students at School 1 used similar reasoning to come to three different conclusions: 
\begin{align}
P_A(1)\gg P_A(2)>P_A(3)&\rightarrow\frac{P_A(3)}{P_A(2)}>\frac{P_A(2)}{P_A(1)};\label{ndd1}\\
P_A(1)>P_A(2)\gg P_A(3)&\rightarrow\frac{P_A(3)}{P_A(2)}<\frac{P_A(2)}{P_A(1)};\label{badgg}\\
P_A(1)\gg P_A(2)\gg P_A(3)&\rightarrow\frac{P_A(3)}{P_A(2)}=\frac{P_A(2)}{P_A(1)}.\label{ndd3}
\end{align}
Interestingly, this third case was used to justify a correct response. 

In each of these cases, students seem to be considering the probabilities in pairs and using the relative difference between each pair to compare the ratios of the pairs. This is consistent with a strategy for comparing fractions that Smith refers to as Compare Numerator-Denominator Differences (NDD) \cite{Smith1995}. The NDD strategy is categorized by students using the within-fraction difference between the denominator and the numerator as a comparative measure. Examples of the NDD strategy in Smith's study include students determining that $3/5=5/7$ (because $5-3=7-5=2$) and that $14/24>7/12$ (because $24-14>12-7$) \cite{Smith1995}.  In our case, students are explicitly or implicitly using the differences between the probabilities as justification for comparing their ratios. Arons cites difficulties interpreting ratios as one of the most prevalent cognitive gaps for students at the secondary and undergraduate levels \cite[p.\ 4--9]{Arons1997}.

Students who ranked the probabilities as simply $P_A(1)>P_A(2)>P_A(3)$ also made claims that are consistent with some of Smith's other classifications. Using this ranking to claim that $P_A(3)/P_A(2)>P_A(2)/P_A(1)$ (``greater than'' on part A) is consistent with Smith's Denominator Principle (DP, fractions with larger denominators are smaller than fractions with smaller denominators), and using this ranking to claim that $P_A(3)/P_A(2)<P_A(2)/P_A(1)$ (``less than'' on part A) is consistent with both the Numerator Principle (NP, fractions with larger numerators are larger) and Larger Components (LC, fractions with larger numerators and denominators are larger) \cite{Smith1995}. However, since no student admitted to exclusively using either the numerator or the denominator of each ratio to compare the two, we cannot be certain that students used these strategies, only that the students' final responses are consistent with their use. 

Students who used probability rankings to justify their answers were categorized as being consistent with one (or more) of Smith's strategies. Reliability of the categories for our classification was checked by an independent classification of the data from School 2. There was initial agreement for 72\% of the student responses; after discussion and negotiation, agreement of 91\% and at least partial agreement of 97\% of students was obtained (one analysis placed some students simultaneously in two categories, while the other only agreed on one of the categories for this group). 

At both School 1 and School 2 more student responses were aligned with the NDD strategy than either DP or NP/LC, and no significant differences were found between the two student populations in terms of their use of these strategies. In many cases it is unclear precisely why a student chose the response s/he did based on the ranking provided, but it is interesting to note the similarities between their claims and those made by the adolescent students in Smith's study.

The key difficulty identified so far is that many students do not apply the \BF\ when it is appropriate to do so, even after lecture instruction. Instead, these students provide responses that are consistent with using novice-like reasoning strategies for comparing ratios. Most students recognized that lower energies are more probable, but they offered no physical justification for why this is so and couldn't use this information alone to make conclusions about the probability ratios in question. 

\subsection{Recitation of a mathematical derivation without physical understanding}
\label{sec:memorize}
In an effort to probe student understanding of the \BF\ more deeply, we conducted individual interviews with four students at School 1 after classroom instruction in the first year of tutorial implementation to determine their familiarity with the \BF, its applications, and its origin.  Two interview participants had participated in the first half of the \bft\ during class (in which they discussed the definitions of macrostates, microstates, and multiplicity for the microcanonical and canonical ensembles, see Sec.\ \ref{sec:bf-tut}), while the other two had not seen the tutorial. The interviews were conducted in the style of a teaching experiment \cite{Steffe2000,Engelhardt2004,Smith2013} and consisted of asking students to complete a guided-inquiry activity that started with asking them to consider how probability relates to multiplicity in the divided container ($C$-$R\,$) scenario (see Fig.\ \ref{sys-res}) and culminated with the derivation of the \BF\ \footnote{Teaching interviews were not audio or video recorded so as to provide a more informal atmosphere. Analysis is based on interviewer field notes and students' written work.}. 

The teaching experiment is a unique form of interview as ``it is an acceptable outcome\dots for students to modify their thinking'' during the course of the interview \cite{Engelhardt2004}. According to Steffe and Thompson, ``a teaching experiment involves a sequence of teaching episodes\dots [including] a teaching agent, one or more students, a witness of the teaching episodes, and a method of recording what transpires during the episode'' \cite{Steffe2000}. For our purposes the interviewer alternated roles as both teaching agent and witness during each interview. In a sense, the activities used during the interview may also be seen as a teaching agent as they included tasks for students to complete, and students interacted with the document in an intellectual manner. Our goal for the interviews was not to simply determine students' understanding of the \BF, but rather to examine how well they could complete instructional tasks based on previous knowledge related to the \BF. Students worked on their own; the interviewer solicited explanations for their work and gave assistance when required. Field notes were taken during the interviews, and students' written work was collected afterward. 

\begin{table}[tbh]
\caption{Sample energy \& multiplicity values for the ``toy model'' system ($C$) and reservoir ($R$; see Fig.\ \ref{sys-res}). This table was presented to students during the teaching interviews and is also used in the \bft. A key element of this situation is that the combined energy of $C$ and $R$ is a fixed value, $E_{tot}$.}
\begin{tabular}{cccc}
\\
\hline\hline
$E_C$&$\omega_C$&$E_R$&$\omega_R$\\
\hline
$E_1$&1&$E_{tot}-E_1$&$3\times 10^{18}$\\
$E_2$&1&$E_{tot}-E_2$&$5\times 10^{19}$\\
$E_3$&1&$E_{tot}-E_3$&$4\times 10^{17}$\\
$E_4$&1&$E_{tot}-E_4$&$1\times 10^{20}$\\
$E_5$&1&$E_{tot}-E_5$&$7\times 10^{18}$\\
\hline
\end{tabular}
\label{mult_tab}
\end{table}%

Results from the teaching interviews provide further evidence of the need for the \bft, especially with regard to the origin of the \BF\ itself. None of the interview participants found the tasks to be trivial, and none correctly articulated how the \BF\ as an expression of probability relates to multiplicity prior to the interview. A major finding during these interviews was the identification of students' difficulties in executing the Taylor series expansion as part of the derivation of the \BF; we reported these difficulties previously \cite{Smith2013}. 

One episode during one of the student interviews was of particular interest.  One student (Joel \footnote{All names are pseudonyms.}, who had participated in portions of the \bft\ in class) was very familiar with the applications of the \BF\ and seemed to be just as familiar with its origin. In one portion of the activity, students were given a table of multiplicities for various discrete system energy levels and asked to determine the most probable macrostate (see Table \ref{mult_tab}). The desired result was for students to conclude that the macrostate with the greatest reservoir multiplicity would be the most probable. Joel wanted to use the \BF\ rather than thinking about multiplicities, even though no information had been given about the relative energy values \footnote{Joel had also provided this reasoning during the in-class tutorial session.}. The interviewer asked Joel to show where the \BF\ came from before applying it to this situation, at which point Joel quoted the textbook derivation of the \BF\ practically verbatim. The final portion of Baierlein's mathematical derivation is as follows \cite[p.\ 92]{Baierlein1999}\footnote{Eq.\ \eqref{joel_eq3} is not explicitly shown in Ref.\ \cite{Baierlein1999}, but Joel wrote it during his interview.},
\begin{align}
P(\psi_j)&=\textrm{const}\times\left(\parbox{1.75in}{multiplicity of reservoir when it has energy $E_{tot}-E_j$}\right)\label{joel_eq1}\\
P(\psi_j)&=\textrm{const}\times\exp\left[\frac{1}{k}S_R(E_{tot}-E_j)\right]\label{joel_eq2}\\
P(\psi_j)&=\textrm{const}\times\exp\frac{1}{k}\left[S_R(E_{tot})\!+\!\frac{\partial S_R}{\partial E_R}\eval{E_{_{tot}}}\!\!\!\!\left(-E_j\right)\right]\label{joel_eq3}\\
P(\psi_j)&=\textrm{(new constant)}\times\exp\left(-E_j/kT\right).\label{joel_eq4}
\end{align}
This derivation exploits the fact that the combined energy of the system and reservoir ($E_{tot}$) is a fixed quantity in order to write the energy of the reservoir ($E_R$) in terms of the energy of the system ($E_j$).

Joel's ability to reproduce the derivation might suggest an understanding of the physical significance of the \BF. However, when asked how the multiplicity of the reservoir relates to the \BF, Joel was at a loss. During his replication of the derivation of the \BF\ he had implicitly written that it was proportional to $\omega_R$ (connecting Eqs. \ref{joel_eq1} and \ref{joel_eq4}), but without explicit help from the interviewer, Joel could not recognize that the multiplicity of the reservoir when it has energy, $E_{tot}-E_j$ (RHS of Eq.\ \eqref{joel_eq1}), is proportional to the exponential function, $\exp\left(-E_j/kT\right)$ (RHS of Eq.\ \eqref{joel_eq4}). Furthermore, Joel had great difficulty relating the physical example used in the textbook (a ``bit of cerium magnesium nitrate\dots in good thermal contact with a relatively large copper disc'' \cite[p.\ 91]{Baierlein1999}) to the ideal gas example used during our interview. He was unable to recognize and articulate the important physical characteristics of each scenario that make the \BF\ applicable; e.g., a system with fixed temperature and variable energy. Joel's failure to make these connections suggests an incomplete understanding of the physical reasoning used to derive the \BF, even after memorizing the textbook derivation.

Results from the teaching interviews and the PRQ suggest that many students can neither recognize situations in which the \BF\ is applicable, nor articulate the physical significance of the \BF\ as an expression for multiplicity, one of the fundamental quantities of statistical mechanics. These difficulties prompted our development of the \bft\ to help students better understand the physical origin of the \BF\ and how it may be applied in various contexts.

\section{Design and Implementation of the \bft}
\label{sec:bf-tut}
Given students' apparent lack of recognition of when to apply the \BF\ to a physical scenario, we designed a guided-inquiry tutorial activity to lead students through its derivation and encourage deep cognitive connections between the physical quantities involved. The derivation chosen for use in the \bft\ is included in the Appendix and may be found in many widely used textbooks, including the one used at the primary research site \cite{Baierlein1999}. 

Our \bft\ gives students the opportunity to productively struggle with the connections between the mathematical formalism and the physical interpretations within the derivation of the \BF\ \cite{Smith2013}. Fostering physics-mathematics connections, such as gaining facility with taking limits and making approximations, as well as knowing when to take these steps, is an important and nontrivial component of upper-division courses as students transition from novices to experts in the field \cite{Bing2012}.

The desired student outcomes during the tutorial are consistent with the concept of \textit{productive disciplinary engagement} (PDE) \cite{Engle2002}.  The small group setting, with explicit instructions to discuss responses and reasoning with group mates, fosters engagement, which is evident through student-student discourse.  Because disciplinary content is the core of the tutorial, most engagement in a tutorial constitutes disciplinary engagement. Finally, Engle and Conant define productive disciplinary engagement as episodes where student are making progress in their engagement with the content \cite{Engle2002}.  Evidence of this productivity includes students recognizing their confusion about a concept or making a new connection as a result of the interaction.  Indeed, the tutorial pedagogy and the materials themselves provide a setting designed to foster PDE.  The 
pedagogy used in the canonical set of tutorials, \textit{Tutorials in Introductory Physics}, to address specific student difficulties, fully matches the parameters of PDE \cite{McDermott2001}\cite[p.\ iii]{McDermott2002}. 
Our primary goal is that students discuss topics in a way that helps them progress through the tutorial tasks while gaining a better understanding of those topics (discussing relevant concepts, synthesizing information, engaging with the connections between the mathematics and the physics, etc.). While in some cases other, less time-consuming pedagogical approaches may also foster PDE and help students with specific difficulties, in this case the depth of the content and the difficulty of the sequence of steps in the derivation of the \BF\ suggested that a tutorial would be the most effective way to achieve this.  Given that a lecture on this topic typically uses an entire class period, we expected the tutorial would occupy a full class period as well as some time outside of class.

\subsection{The \textit{Boltzmann Factor} Tutorial}
The \bft\ begins by asking students to consider an isolated container of an ideal gas. They are guided to recognize that the container has a fixed internal energy ($E_{tot}$) and that all accessible microstates are equally probable.

Once the properties of the contents of the isolated container have been established, the students are presented with a scenario in which the container of ideal gas is separated into relatively small and large sections (see Fig.\ \ref{sys-res}). The small system of interest ($C\,$) is said to be in thermal equilibrium with the large reservoir ($R\,$), and the students are asked to compare the values of various thermodynamic properties of $C$ to those of $R$ to highlight the fact that the \textit{intensive} properties (temperature, pressure) will have the same value for both $C$ and $R$, while the values of the \textit{extensive} properties (volume, number of particles, internal energy) of $C$ are much smaller than those of $R$.  

The third section of the tutorial uses the fact that the multiplicities of $C$ and $R$ are so different ($\omega_C\nobreak\ll\nobreak\omega_R$) to justify a single-particle ``toy model'' in which $\omega_C\nobreak=\nobreak1$ ($\omega_{tot}\nobreak=\nobreak\omega_C\,\omega_R\nobreak=\nobreak\omega_R$), and the energy of $C$ can only take on a handful of discrete values, $E_C\nobreak\in\nobreak\{E_j\}\nobreak=\nobreak\{E_1,E_2,\dots\}$ (see Table \ref{mult_tab}).  The students are asked to determine which system microstate ($j=1$, 2, 3, 4, or 5) is most probable and which is least probable. The desired solution is that the system microstate in which the macrostate of $R$ has the largest multiplicity ($\omega_R$) is the most probable ($E_4$ in Table \ref{mult_tab}) because all reservoir microstates are equally likely. Careful consideration of the relative probabilities of each macrostate leads to the proportionality between the probability of the $j^\mathrm{th}$ microstate of $C$ and the multiplicity of the reservoir: $P(\psi_j)\nobreak\propto\nobreak\omega_R(\psi_j)$.

The final section of the \bft\ is the derivation of the \BF\ itself. The core of this derivation is a Taylor series expansion of $S_R(E_R)$ about  the value $E_R\nobreak=E_{tot}$ to obtain the expression for $S_R$ as a linear function of $E_C$ given in Eq.\ \eqref{taylor} \cite{Smith2013}. The students are explicitly asked to consider the physical significance of each term in the expansion and to determine the final linear expression on their own. Then, using the relationship between entropy and multiplicity in Eq.\ \eqref{S}, they are guided to derive an expression for $\omega_R$:
\begin{equation}\omega_R=e^{S_R/k}=e^{S_R(E_{tot})/k-E_C/kT},
\end{equation}
and because $S_R(E_{tot})$ is a constant,
\begin{equation}\omega_R\propto e^{-E_C/kT},\end{equation} 
i.e., the \BF. Students find that $P(\psi_j)\nobreak\propto\nobreak\omega_R(\psi_j)$ and that $\omega_R(\psi_j)\nobreak\propto e^{-E_j/kT}$, leading to the proportionality in Eq.\ \eqref{bf}. Finally, they obtain the expression for the \Z, $Z$, by normalizing the probability. 

The post-tutorial homework assignment is an application of the Boltzmann factor to a three-state system with unevenly spaced energy levels. Students are asked various questions about the ratios of probabilities of the system being in a particular state. These questions are similar to the PRQ but the students are given specific values for $T$ and $N$ and asked to determine numerical values for the probability ratio rather than compare two different ratios. They are also asked to determine an expression for the generic ratio between the probabilities of any two energy levels. This homework assignment was used as a continuation of the tutorial, not as an assessment or research tool. 

\subsection{Tutorial Implementation}
\label{sec:bf-implementation}
At School 1, the \bft\ was implemented after all lecture instruction on the \BF. Students were given one 50-minute class period to complete the tutorial. The course instructor and one additional facilitator were available during the tutorial session as observers and facilitators \footnote{All students at School 1 had participated in similar tutorials in at least one previous course.}. No course credit is offered for participation in the tutorial itself, but the course grade does include a component for class participation. Several groups were videotaped during tutorial sessions (in three years of classes) to monitor tutorial progress and document student reasoning regarding the \BF\ and related topics. 

The \bft\ was implemented at School 2 once in place of lecture instruction. Students were given one 50-minute class period and an additional 20 minutes during the next period to complete the tutorial. As described in Sec.\ \ref{sec:prq-pre}, the PRQ was administered as a pretest at both institutions before tutorial instruction, and a similar question was used on course examinations. 
 
\section{Findings during implementation of the \textit{Boltzmann factor} tutorial}
An interesting question is whether recognizing when to use the \BF\ serves as a direct proxy for an understanding of the physical significance and meaning of the expression --- its origin and why it describes the relative occupation of states.  Instructors typically assume that this is the outcome of presenting the derivation of such functions to students: that the clear description of the steps of the derivation, including the explicit connections between the mathematical steps and the physical constraints, assumptions, etc. that drive the mathematics, provides students with the intended insight.  Thus the subsequent assumption is that the proper invocation of the \BF\ implies an understanding of its meaning and significance.  However, in the process of pilot-testing the \bft\ we observed that this is not necessarily the case. We have evidence from students working through sections of the tutorial either in class or during teaching interviews that suggest that (a) the students do not have a sense of the physical basis for the \BF\ before the tutorial and (b) the sequencing of the tutorial provides the students the opportunity to gain an understanding and appreciation for this physical foundation.

Below we describe the findings from the data collected during and after tutorial implementation.  These data --- collected in written and video form --- provide evidence to support the claim that the activities the students work through in the tutorial improve both students' ability to apply the \BF\ appropriately and their understanding of the physical basis for the \BF, including the connections between some of the mathematical steps and the physical scenario.  

\begin{figure*}[tb]
\begin{center}
\includegraphics[width=6.5in]{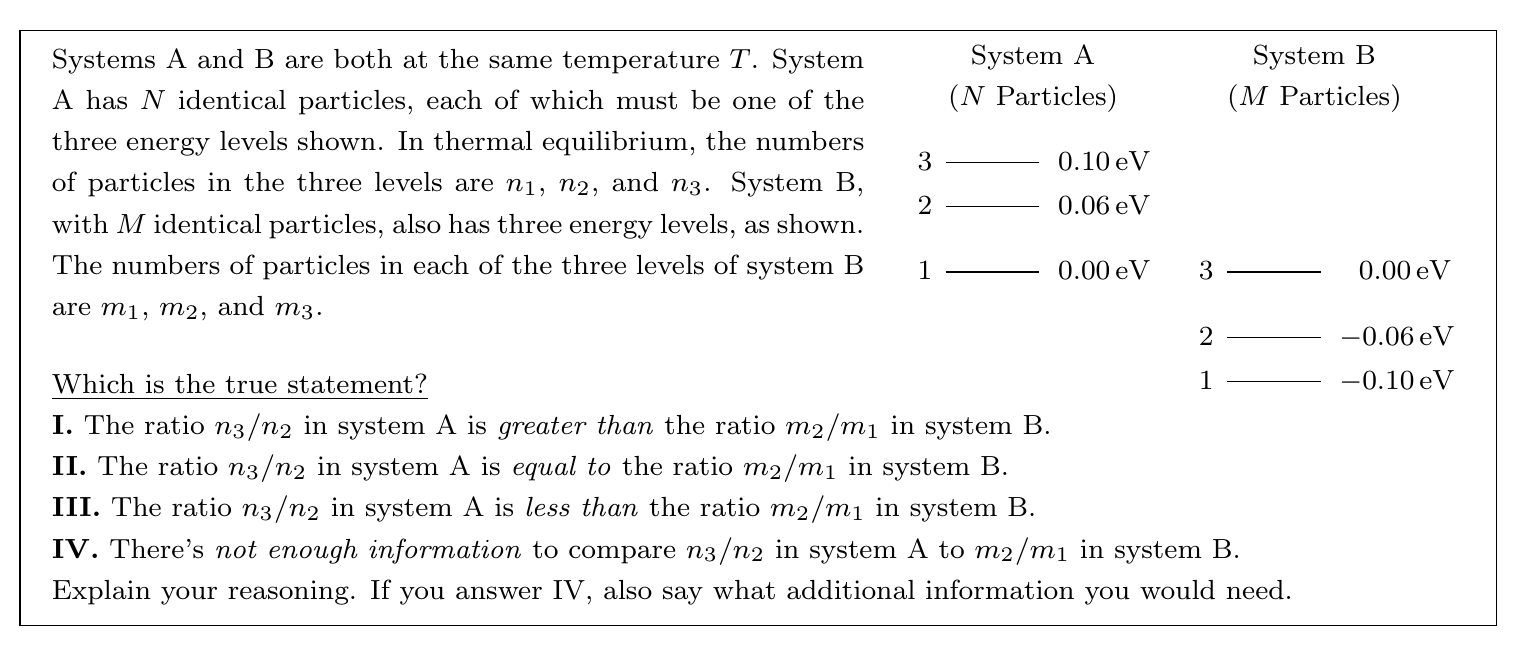}
\end{center}
\caption{PRQ Analog: developed by instructor at School 2. Administered on a course exam once at School 1 and at School 2.}
\label{knight-prq}
\end{figure*}

\subsection{Improving student use of the \BF\ in appropriate contexts}
\label{sec:bf-exam}
\begin{figure}[tb]
\begin{center}
\includegraphics[width=3.4in]{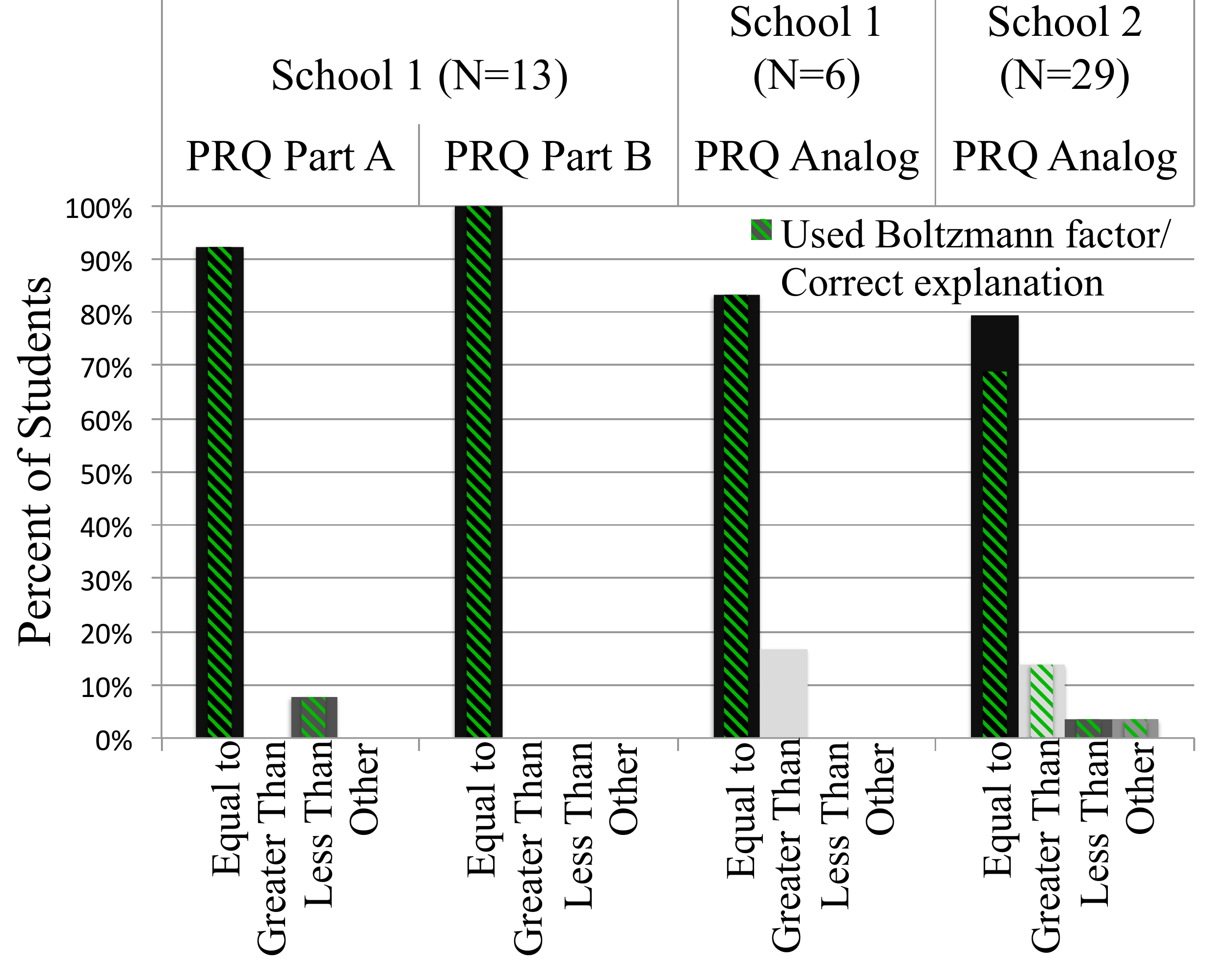}
\end{center}
\caption{Post-tutorial results from PRQ and PRQ Analog, administered during course examinations. The green diagonal stripes indicate the students who used the Boltzmann factor or stated that the energy of the ground state was irrelevant (in part B) to obtain their chosen answer(s). For the PRQ Analog, ``Equal to'' corresponds to choice II, ``Greater than'' to choice I, ``Less than'' to choice III, and ``Other'' to choice IV. Data are shown for students who completed the PRQ pretest, participated in the \bft, and completed either the PRQ or PRQ Analog exam question.}
\label{prq-post}
\end{figure}

In order to probe the effect of tutorial instruction on student tendency to invoke the \BF\ in an appropriate situation, we administered written post-tests at both schools on midterm examinations.  The PRQ was given on a course examination after the \bft\ in two years at School 1. A similar question, referred to as the PRQ Analog (shown in Fig.\ \ref{knight-prq}), was developed by the instructor at School 2 and asked on a course exam in one year at both institutions. The PRQ Analog requires students to apply the same knowledge as is used to correctly answer the PRQ: that the probability of a particle being in one of the energy states is proportional to the \BF, and that a ratio of probabilities would be equivalent to a ratio of \BF s, which depends only on the difference between the energy levels. However, the PRQ Analog adds some complexity by using systems with energy levels that are not evenly spaced and requiring students to recognize that the number of particles occupying each energy level will be proportional to the probability of a single particle having that energy. Despite this added complexity, Fig.\ \ref{prq-post} shows that students' use of the \BF\ was very similar on both the PRQ and PRQ Analog exam questions.

From the three implementations at School 1 there are 19 sets of matched (pre-/post-tutorial) data of students who participated in the \bft. There are 29 sets of matched data from School 2. Figure \ref{prq-post} shows the exam data from all students broken down by question and school. These data provide evidence that the \bft\ helps students recognize the utility of the \BF\ and how to apply it properly in the context of these questions.

The most striking feature of Fig.\ \ref{prq-post} is that all 13 students at School 1 used appropriate \BF\ reasoning on both parts of the PRQ \footnote{The single student who answered ``less than'' on part (a) of the PRQ correctly used the Boltzmann factor to find the probability of each energy level but made errors when calculating the ratios.}. Moreover, all but one student at School 1 (about 83\%) used the \BF\ correctly to answer the PRQ Analog after tutorial instruction. This is a marked improvement over lecture instruction alone (only about half consistently used the \BF\ on the PRQ pretest). Similarly, all but two students at School 2 (also almost 95\%) used the \BF\ to answer the PRQ Analog exam question.  

In order to perform statistical analyses to compare the exam results with the pretest results, data were grouped into the four categories discussed in section \ref{sec:pre-bf}. This reduced coding scheme is necessary because we essentially asked three questions at various times (PRQ parts A and B, and the PRQ Analog): the specific responses to the various questions, e.g., ``greater than,'' cannot necessarily be considered the same response. As such, the only categories available for grouping responses are either the correct response or one of the incorrect responses; along with this we have the dimension of whether or not a student used the \BF\ appropriately to justify his or her response, consistent with the four categories above. These general categories do not allow claims to be made about how reasoning patterns differ within the incorrect responses, but they do allow comparisons of the frequency with which students use the correct \BF\ reasoning and whether or not it yielded a correct response. Using these categories, a Fisher's exact test showed that all exam data are statistically similar ($p=0.125$).  Additionally, a Fisher's exact test comparing the data at School 1 showed that the results from the exams are statistically significantly better than the results on the PRQ pretest on both parts A ($p\nobreak=\nobreak0.019$) and B ($p\nobreak=\nobreak0.012$) \footnote{Fisher's exact tests comparing pre tutorial and posttutorial results from School 1 were conducted using only matched data from students who had participated in our \bft. In general these students performed better on the PRQ pretest than the overall data set shown in Fig.\ \ref{bf-pre} (47\% use the \BF\ to obtain the correct answer to part A as opposed to 34\%), but there is still a significant difference in their performance after participating in the tutorial.}. A Fisher's exact test also shows that the exam results at School 2 are significantly better than the pretest data ($p<0.001$ for both parts). 

The written pretest results suggested that students were not aware of the contexts in which the Boltzmann factor is applicable, even after lecture instruction.  The written post-test results demonstrate a marked improvement in the correct use of the \BF\ in these situations. These results suggest that the \bft\ helps improve student understanding of how and when to use the \BF\ when it is used either as a stand-alone activity (School 2) or as a supplement to lecture instruction (School 1).

\subsection{Improving student understanding of the physical basis for the Boltzmann factor}
\label{sec:bf-meaning}

As mentioned above, in addition to improving appropriate student use of the \BF, the other major goal for the \bft\ was for students to gain an appreciation for the physical basis of the \BF, which did not occur based on lecture instruction, even when a student was able to recite the derivation exactly, as shown in section \ref{sec:memorize}.  However, we anticipated that students would gain this appreciation by working through the \BF\ derivation in small groups while emphasizing the physical justifications for each step therein. Documenting the acquisition of this appreciation or understanding is not possible using written data of the sort typically gathered. So, in order to monitor student progress and success in achieving this instructional goal for the \bft, we videotaped several groups of students while they completed the tutorial during the first three years of implementation at School 1. 

Segments from these classroom episodes were selected for transcription and further analysis based on the content of student discussions. Given our focus on investigating students' understanding of particular topics, our methods of gathering video data align with Erickson's description of \textit{manifest content} approaches, in which particular classroom sessions are selected to be videotaped based on the content being discussed \cite{Erickson2006}. We chose to videotape classroom sessions in which students were engaging in our tutorial because we are primarily interested in their ideas regarding the conceptual and mathematical content of our tutorial and students' ability to negotiate tutorial prompts in an efficient and productive manner.  We have already explained that our use of ``productive'' follows its use in \textit{productive disciplinary engagement} \cite{Engle2002}; we classify ``efficient'' interactions as those enabling the students to complete the tutorial within the intended 50-minute class period. In some respects this categorization of student interactions is done with an eye toward the end justifying the means: an interaction cannot necessarily be considered productive or efficient without knowing the conversations that take place after that interaction.

Over three years of tutorial implementation we videotaped a total of four groups containing 13 students. To analyze video data, we watched each video in its entirety and made note of conversations that seemed interesting; we later watched these segments many times and recorded both what was discussed and why we thought it was interesting. Quotations included in this section were often selected for their uniqueness. Several students made comments and statements that indicated difficulties that were not expected and have not been previously documented. Data do not exist to verify the pervasiveness of these difficulties, but we feel their existence is noteworthy. In cases where more than one student displayed a similar difficulty, we have included multiple quotes to allow the reader to evaluate the similarities and differences between the data.

Video data from the second tutorial implementation at School 1 provide evidence that students gain an appreciation for the origin of the \BF\ while participating in the \bft. Two students (Sam and Bill, who worked in a group on their own) participated in several conversations throughout the tutorial session that  indicate their contemplation of relevant physical ideas. During the \bft\ they discussed which macrostate (from Table \ref{mult_tab}) is most probable:

\begin{tabular}{r@{~~--~~}p{6.25cm}}
\multicolumn{2}{c}{~}\\
Bill&Probably the one with more microstates\\
Sam&Yeah\dots the one with the highest multiplicity\\
\multicolumn{2}{l}{\hspace{10mm}\dots}\\
Bill&``Give a general expression for the probability of the system''\dots so probably just use omega R ($\omega_R$), so we'd say omega R j ($\omega_{R_j}$) over the sum of all of them.\\
Sam&Yeah, that's what we said: omega R j over the sum of omega R j $\left(\omega_{R_j}\big /\sum\omega_{R_j}\right)$.\\
\multicolumn{2}{c}{~}\\
\end{tabular}

Later in the tutorial, after completing the Taylor series expansion (with instructor intervention), interpreting the physical quantities involved, and relating their expression for multiplicity to the Taylor series of entropy, Sam and Bill had a realization \footnote{See Ref.\ \cite{Smith2013} for more details on these students' successes and difficulties with the Taylor series portion of the tutorial.}:

\begin{tabular}{r@{~~--~~}p{6.25cm}}
\multicolumn{2}{c}{~}\\
Sam&That's cool. Look, see, you get the Boltzmann factor. You solve for omega ($\omega$): e to the minus E over k T ($e^{-E/kT}$).\\
\multicolumn{2}{l}{\hspace{10mm}\dots}\\
Bill&I guess that's where it comes from.\\
Sam&'Cause we didn't know where it came from.\\
Bill&I had no idea.\\
Sam&I was just like, ``OK.''\\
\multicolumn{2}{c}{~}\\
\end{tabular}

These excerpts indicate that Sam and Bill are discussing relevant physical quantities and principles and gaining an appreciation for the origin of the \BF\ as a result of the \bft. In particular they are correctly relating the \BF\ of the system with the multiplicity of the reservoir as an indicator of probability. It should be noted that before tutorial instruction, Sam answered both parts of the PRQ correctly using correct reasoning, and Bill used the \BF\ correctly but made errors in his calculations. These data indicate that students who are able to successfully use the \BF\ after lecture instruction may not have a complete understanding of the conceptual meaning behind the mathematics they are using.

During the third year of tutorial implementation one group struggled to interpret the derivative of entropy (with respect to energy) obtained from completing the Taylor series as the inverse of the temperature $T^{-1}$ (see Ref.\ \cite{Smith2013} for details). However, once they had written an expression for the entropy of the reservoir, one student had a particularly expressive realization upon solving for the multiplicity, 
\begin{quote}
``Actually wait, ohhh, heyyy, because then that becomes the partition [function]\dots and there's your Boltzmann factor.''
\end{quote} 
Similar statements were made by Jake\label{jake}(who had participated in the first three sections of the tutorial in class) and others during the teaching interviews (see Sec.\ \ref{sec:memorize}), indicating that they had not developed a robust understanding of the physical significance of the \BF\ after lecture instruction alone. All observation and interview data indicate that these same students can gain an appreciation for the physical significance of the \BF\ while participating in the \bft.

\subsection{Revising the tutorial}
\label{sec:bf-revisions}

The development process for instructional materials is iterative; modifications are typically made to improve the instructional experience based on earlier implementation(s). For this reason data are collected during the tutorial implementations to ascertain the impact the materials are having on students' abilities to interact with the tutorial activities, including the extent to which: (a) students interpret the instructions as the developers intended; (b) the tasks and questions in the tutorial generate productive discussions among the students, elicit specific difficulties targeted by the developers, and guide students to the desired outcomes; and (c) students are able to complete the tutorial tasks in the allotted time. The video data from School 1 serve this purpose, as does detailed feedback from the instructor at School 2 regarding students' abilities to perform tutorial tasks as well as specific places where they had particular difficulty. Additional data came from the teaching interviews described in section \ref{sec:memorize}, which were conducted after the initial implementation at School 1 (when many students did not complete the tutorial). In this authentic instructional setting we find evidence of additional specific difficulties that written data would not elicit, as well as examples of student discussions prompted by the materials that inform the development of the tutorial and further research.

All of these data were used to inform tutorial revisions and modifications. Some revisions were minor, such as wording changes to improve clarity for the students.  Other changes were more extensive, and included removing sections entirely or moving activities and tasks to be completed either as pre-tutorial homework (the initial steps in the Taylor series expansion of $S_R(E_R)$ \cite{Smith2013}) or post-tutorial homework (obtaining the expression for the \Z).  It should be noted that data do not exist to determine the precise effect that each individual tutorial modification has on student learning and understanding of the \BF. However, the data do suggest that the collective modifications have led to increased student efficiency in completing tutorial tasks during later implementations, allowing students to complete more of the tutorial in the time allotted. Increased efficiency benefits students by giving them the opportunity to arrive at the ``punchline'' of the \bft: the derivation of the \BF\ itself.

During the first tutorial implementation at School 1 several unanticipated difficulties were observed. The first occurred while students completed the first page of the tutorial on which it asked them to ``estimate (to order of magnitude) how many \textit{microstates} (molecular configurations) exist such that the total energy of the gas [in the isolated container] is $E_{tot}$.'' This language cued the students to attempt to find a formula for calculating the multiplicity of the gas based on its energy \footnote{Discussions centered around trying to remember the density of states function and the binomial distribution.}. The intent of the task, however, was for the students to recognize that there would be many many molecular configurations that would have a total energy of $E_{tot}$ and to just write down any appropriately large number. Students spent four minutes on this task before asking the instructor for help. (This wasn't expected to take very long; a rigorous calculation was neither intended nor possible, and thus it should only have taken about a minute.) The wording of the question was altered in subsequent implementations to ask the students, ``How many \textit{microstates} (molecular configurations) would you estimate exist such that the total energy of the gas is $E_{tot}$: 1, 1000, $10^N$?'' Data from the second tutorial implementation at School 1 indicate that students found this order-of-magnitude estimate much easier than the year before.

One observation noted during the teaching interviews was that some students focused strongly on a relationship between multiplicity and energy $\left(\omega\nobreak\propto V^NE^{3N/2+1}\right)$ that was given in an introductory paragraph of the interview (and the tutorial section). The intent of the statement was to connect the \bft\ to the density of states function $\left(D(E)\equiv\D\omega/\D E\nobreak\propto V^NE^{3N/2}\right)$, which they had recently learned about, and to motivate the notion that $\omega_C\nobreak\ll\nobreak\omega_R$ (given that $V_C\nobreak\ll V_R$ and $E_C\nobreak\ll E_R$). However, students tried to use this expression to relate the multiplicities given in Table \ref{mult_tab} to the energies. One student (Jake, see p.\ \pageref{jake}) even stated that since the $E_C=E_3$ microstate has the lowest reservoir multiplicity ($\omega_R\nobreak=\nobreak4\nobreak\times\nobreak10^{17}$, rightmost column in Table \ref{mult_tab}), $E_3$ must be the lowest energy (of $C\,$) and, therefore, be the most probable. What he failed to consider is that the multiplicity of the \textit{reservoir} is the lowest, making $E_R$ the lowest, and $E_3$ the highest value (by conservation of energy). Jake's reasoning, in fact, reached the exact opposite conclusion of what was intended. 

The intent of the energy/multiplicity table (Table \ref{mult_tab}) and related questions is to motivate the connection between multiplicity of the reservoir and probability of the system being in the corresponding microstate. The students were meant to realize that the $E_C=E_4$ microstate is the most probable since it has the largest corresponding multiplicity for the reservoir, leading them to conclude that $E_4$ must be the lowest energy of the system because $E_R$ must be at its highest value. Two other interview participants displayed this tendency to latch onto the given expression relating multiplicity to energy; it was also observed during the in-class tutorial session to a lesser extent. The statement reminding students about the connection between multiplicity and energy was removed from later implementations of the \bft\ along with most of the original introductory paragraph. 

Other in-class observations indicated that students did not always refer to their own work from previous sections of the tutorial when answering more difficult questions later. In particular, when answering questions about multiplicity concerning the divided container (see Fig.\ \ref{sys-res}), students did not necessarily refer to the conclusions they had made about the original undivided container. Specific references to previous tutorial sections were added to encourage students to make these connections and build on knowledge they had previously constructed. 

The most consistent observation from the first two implementations of the \bft\ at School 1 and the implementation at School 2 is that students could not complete the tutorial in one 50-minute class session. The students at School 1 during the first year were only able to complete the first three sections of the tutorial, ending in an expression indicating $P(\psi_j)\nobreak\propto\nobreak\omega_R(\psi_j)$. They did not have the opportunity to even begin the Taylor series expansion that would lead to the derivation of the \BF\ (the portion of the tutorial that we expected to be the most difficult). After revising the tutorial to address the specific difficulties discussed above, students at School 1 were able to successfully complete the first four sections of the tutorial (culminating with the derivation of the \BF) within one 50-minute class during the second year, but they were not able to complete the normalization of probability to determine an expression for the \Z. A similar result was reported at School 2 in that six out of seven groups of students ($\approx4$ students per group) were able to derive the \BF\ after the entire 70 minutes allotted by the instructor, but only 1-2 groups had enough time to derive $Z$ as well. The students who did work through that portion of the tutorial, both those in class at School 2 and those in the teaching interviews at School 1, had little trouble normalizing their expression for probability to get $Z$. 

Based on the overwhelming majority of students not completing the entire tutorial, even after modifications, we removed the fifth section of the tutorial, in which students derive the \Z\ from the in-class activities, and added it as the first question in the post-tutorial homework assignment. The in-class portion of the tutorial now ends with the derivation of the \BF\ as well as a comment on the term ``\BF'' and a reference to the homework assignment in which students will determine an exact expression for the probability rather than just a proportionality. Classroom observations from subsequent implementations at School 1 indicate that these revisions have improved the efficiency of the tutorial, and that most students are able to complete the derivation of the \BF\ during a single 50-minute class period \footnote{Results from later implementations are from instructor notes. No video data exist after the second implementation at School 1.}.

\section{Summary and Implications for Future Work}
\label{sec:bf-conc}

Our results show that students often do not use the \BF\ when answering questions related to probability in applicable physical situations after lecture instruction alone. These results have been replicated over several years. Students instead tend to use statements about a ranking of the relative probabilities to make novice-like claims about probability ratios, consistent with literature in mathematics education. This is a common error among students regardless of whether or not they had received lecture instruction on the \BF. To address students' failure to appropriately apply the \BF, we developed the \bft\ to improve their understanding of situations in which the \BF\ is appropriate by guiding them through a derivation of the \BF, one that is particularly rich in connecting the physics to the progression through the derivation.  Modifications were made to the tutorial based on teaching interviews and in-class observations in order to optimize student productive disciplinary engagement during class time. Results from several tutorial implementations indicate that students are far more likely to use the \BF\ properly after tutorial instruction than after lecture instruction alone (results are statistically significant at the $p<0.05$ level). The \bft\ can be an effective supplement to (as at School 1) or replacement for (as at School 2) lecture instruction. 

We anticipated that guiding students through this particular derivation of the \BF\ would provide them with the opportunity to engage in the physical reasoning behind the derivation of the \BF, which our data suggested was not an outcome of lecture instruction --- even for a student who invests the effort to memorize the textbook derivation. We have shown that participating in tutorial instruction on this derivation helps students gain an appreciation of the physical implications and meaning of the mathematical formalism behind the formula that had previously eluded them; e.g., Sam and Bill.  

We have previously reported two related studies on students' understanding of Taylor series expansions \cite{Smith2013} and the relationship between the \BF\ and the density of states as expressions of multiplicity \cite{Smith2013a}. These results support our current claim that deriving the Boltzmann factor is subtle and complex, and a robust understanding of its physical meaning is not trivial.

One major avenue for future research is a study on the pervasiveness of student understanding of the \BF\ after tutorial instruction. Do students use the Boltzmann factor appropriately in situations that do not involve probability ratios of discrete, non-degenerate energy states? Do they recognize situations in which the \BF\ is and is not applicable? Our original study has been necessarily focused on helping students understand a basic application of the Boltzmann factor. However, the Boltzmann factor is considered ``the most powerful tool in all of statistical mechanics" \cite{Schroeder2000}. Do students understand this tool well enough to use it to maximum potential? Studying how students use the \BF\ and the \Z\ to derive other physical quantities and investigating how well students understand the physical significance behind the relevant mathematical procedures could help answer this question and provide better insight into what students do and do not understand about the ``quintessential expression of the statistical mechanical approach" \cite{Baierlein1999}.

\begin{acknowledgments}
We thank members of the University of Maine Physics Education Research Laboratory, and our other colleagues, Warren Christensen, Michael Loverude and David Meltzer, for their continued collaboration and feedback on this work. We especially thank Jessica Clark for her assistance with data analysis. We are deeply indebted to the instructors of the courses in which data were collected. This material is based upon work supported by the National Science Foundation under Grant Nos.\ PHY-0406764 and DUE-0817282. 
\end{acknowledgments}

\appendix

\section{Derivation of the \BF}
To understand the mathematical form of the Boltzmann factor, consider the interactions between the system under investigation (we call this $C$ to avoid confusion with entropy, $S\,$) and the thermal reservoir ($R$; see Fig.\ \ref{sys-res}) \footnote{The derivation presented here follows that in many thermal physics textbooks, e.g., Refs.\ \cite{Baierlein1999}, \cite{Reif1965}, \cite{Kittel1980}.}. The probability of finding the system in a particular state will depend on the total multiplicity of the system-reservoir combination ($P(E_C)\nobreak\propto\nobreak\omega_{tot}$), which is the product of the individual multiplicities of the system and the reservoir ($\omega_{tot}\nobreak=\nobreak{\omega_C\,\omega_R}$). In fact, if one considers a small enough system (perhaps a single particle) the energy of the system may only occupy a handful of discrete energy levels ($E_C\nobreak\in\nobreak\{E_j\}\nobreak=\nobreak\{E_1,\,E_2,\,\ldots\}$). If these energy levels are non-degenerate, then the system would have a constant multiplicity, $\omega_C\nobreak=\nobreak1$ \footnote{One may also assume that one is trying to find the probability of the system occupying a particular microstate, so that $\omega_C=1$ even if the energy levels are degenerate.}. The total multiplicity of the system-reservoir combination will then be exactly equal to the multiplicity of the reservoir:
\begin{equation}
\omega_{tot}=\omega_R\textrm{\hspace{0.5mm}}\omega_C=\omega_R.\label{wtot}
\end{equation}
The challenge now is to determine an expression for $\omega_R$ in terms of $E_C$ (the defining parameter of the macrostate). To accomplish this one must first relate $E_C$ to the properties of the reservoir.

It is reasonable to assume that the system-reservoir combination is isolated from the rest of the universe such that its total energy,
\begin{equation}
E_{tot}=E_C+E_R,\label{Etot}
\end{equation}
remains constant. The energy of the system, however, may fluctuate about some average value,
\begin{equation}
E_C=\langle E_C\rangle\pm\delta E.\label{fluctuate}
\end{equation}
The magnitude of these energy fluctuations ($\delta E$) may be relatively large compared to $\langle E_C\rangle$, but insignificant compared to $\langle E_R\rangle$, thus we are justified in considering $R$ a reservoir as its energy does not change appreciably. Qualitatively, by conservation of energy, as the energy of the system decreases, the energy of the reservoir must increase, increasing $\omega_R$ and $\omega_{tot}$, yielding a higher probability; therefore, lower energy states for the system ($C$) are more probable than higher energy states. 

One must now be concerned with the precise mathematical form of multiplicity as it relates to energy, but while energy is an extensive variable, multiplicity is neither extensive nor intensive. This dilemma is solved by relating multiplicity to the extensive quantity entropy via Eq.\ \eqref{S}. Given that entropy is an extensive variable, it may also be written as a function of other extensive variables; e.g., as $S_R(E_R)$. Because the reservoir is so much larger than the system, $E_R=E_{tot}-E_C\approx E_{tot}$, and a Taylor series expansion is appropriate to approximate $S_R(E_R)$ about the point $E_R\nobreak=E_{tot}$:
\begin{align}
S_R(E_R)&=S_R(E_{tot})-\frac{\partial S_R}{\partial E_R}\eval{E_{_{tot}}}E_C+\ldots\nonumber\\
&= S_R(E_{tot})-\frac{E_C}{T},\label{taylor}
\end{align}
where $\left(\partial S\middle/\partial E\right)_{_{_{V,N}}}\textrm{\hspace{-3mm}}=\nobreak T^{-1}$ from the fundamental thermodynamic relation ($\D E=T\,\D S-P\,\D V+\mu\,\D N$) 
and $E_R\nobreak=\nobreak{E_{tot}}\nobreak-\nobreak{E_C}$. The equality in the second line is valid because the temperature of the system (and reservoir) is fixed: higher-order derivatives of entropy are derivatives of temperature and thus vanish. In this manner one obtains an expression for $S_R$ as a function of $E_C$ and constants. Revisiting Eq.\ \eqref{S} one obtains,
\begin{equation}
\omega_R\propto e^{-E_C/kT}\,\therefore\,\,P(E_C)\propto e^{-E_C/kT},\label{wr}
\end{equation}
giving the desired result of $P(E_C)$, from Eq.\ \eqref{bf}.

It should be noted that the above method is not the only way to derive the \BF. Schroeder, for example, uses an approximation of the fundamental thermodynamic relation rather than a Taylor series expansion to determine an expression for $S_R$ in terms of $E_C$ \cite{Schroeder2000}. Carter, on the other hand, uses the method of Lagrange multipliers to maximize $\ln(\omega)$ with the constraints that the average energy and number of particles in the system are both fixed; this derivation does not require the assumption of a large thermal reservoir, as the multiplicity of the reservoir is never used \cite{Carter2001}. The derivation presented in this section was chosen for use within our \bft\ as it is presented in the textbook used at the primary research site \cite{Baierlein1999} as well as several other commonly used texts (see Ref.\ \cite{Reif1965} \& \cite{Kittel1980}) and because the physical significance of the \BF\ (the multiplicity of the reservoir/surroundings) is emphasized throughout.

\bibliography{TIS}

\end{document}